\documentclass[12pt,a4paper,twoside]{article}
\usepackage{epsfig}
\usepackage{amsmath}
\usepackage{latexsym}
\newtheorem{theorem}{Theorem}[section]

\newtheorem{lemma}[theorem]{Lemma}
\newtheorem{example}[theorem]{Example}
\newtheorem{corollary}[theorem]{Corollary}
\newtheorem{remark}[theorem]{Remark}

\newcommand{\pf}{\noindent {\bf Proof: \ }}
\newcommand\Q[1]{$#1$ \quad}

\font\msbm=msbm10 at 12pt
\newcommand{\ZZ}{\mbox{\msbm Z}}
\newcommand{\RR}{\mbox{\msbm R}}

\newcommand{\FF}{\mbox{\msbm F}}

\font\msbb=msbm10 at 22pt
\newcommand{\FFF}{\mbox{\msbb F}}

\begin{document}
\title{Cyclic Codes over the Matrix Ring $M_2(\FFF_p)$\\ and Their Isometric Images over $\FFF_{p^2}+u\FFF_{p^2}$\thanks{This paper was presented at the International Zurich Seminar on Communications, Zurich, Switzerland, February 2014.} } 
\author{Dixie Falcunit, Jr.\\
IMSP, UP Los Ba\~{n}os\\
College 4031, Laguna, Philippines\\
e-mail: {\tt dffalcunitjr@uplb.edu.ph}\\ \\
Virgilio Sison\thanks{This author gratefully acknowledges financial support from Commission on Higher Education, UPLB Academic Development Fund, and UP Research Dissemination Grant.}\\
IMSP, UP Los Ba\~{n}os\\
College 4031, Laguna, Philippines\\
e-mail: {\tt vpsison@up.edu.ph}} 
\maketitle

\begin{abstract}
 Let $\FF_p$ be the prime field with $p$ elements.  We derive the homogeneous weight on the Frobenius matrix ring $M_2(\FF_p)$ in terms of the generating character. We also give a generalization of the Lee weight on the finite chain ring $\FF_{p^2}+u\FF_{p^2}$ where $u^2=0$. A non-commutative ring, denoted by $\mathcal{F}_{p^2}+\mathbf{v}_p \mathcal{F}_{p^2}$, $\mathbf{v}_p$ an involution in $M_2(\FF_p)$, that is isomorphic to $M_2(\FF_p)$ and is a left $\FF_{p^2}$- vector space, is constructed through a unital embedding $\tau$ from $\FF_{p^2}$ to $M_2(\FF_p)$.  The elements of $\mathcal{F}_{p^2}$ come from $M_2(\FF_p)$ such that $\tau(\FF_{p^2})=\mathcal{F}_{p^2}$.  The irreducible polynomial $f(x)=x^2+x+(p-1) \in \FF_p[x]$ required in $\tau$ restricts our study of cyclic codes over $M_2(\FF_p)$ endowed with the Bachoc weight to the case $p\equiv$ $2$ or $3$ mod $5$. The images of these codes via a left $\FF_p$-module isometry are additive cyclic codes over $\FF_{p^2}+u\FF_{p^2}$ endowed with the Lee weight. New examples of such codes are given.

\end{abstract} 


\section{Introduction}
\label{sec:int}
The theory of codes over finite rings has gained much attention since the significant result in \cite{ham:kum:cal:slo:sol} showed that several well-known families of good nonlinear binary codes can be identified as Gray images of linear codes over the quaternary ring $\ZZ_{4}$ of integers modulo $4$. Several recent papers dealt with codes over finite Frobenius rings. These rings are considered the most appropriate coding alphabet since the two classical theorems, namely the extension theorem and the MacWilliams identities, generalize neatly in the case of finite Frobenius rings.

Let $p$ be a prime and $r \ge 1$ an integer. We denote by $\FF_{p^r}$ the Galois field of order $p^r$ and characteristic $p$. In this study we restrict ourselves to a small class of finite Frobenius rings, the matrix rings over a finite field, in particular the ring of $2\times 2$ matrices over $\FF_p$, denoted by $M_2(\FF_p)$. The multiplicative group $GL(2,p)$ of invertible matrices in $M_2(\FF_p)$ will be of much use in the ensuing discussion as well. Until now very few publications on codes over non-commutative rings have been seen. It was only in 2012 that the theory of cyclic codes over $ M_2(\FF_{2})$ was developed \cite{ala:sbo:sol:yem}. The idea for the construction of cyclic codes over $M_2(\FF_{2})$ came from \cite{bac} in which was defined an isometric map $\phi$ from $\FF_4^2$ onto $M_2(\FF_2)$ where
\[ \phi((a+b\omega,c+d\omega))=\begin{pmatrix} a+d & b+c\\
                                                                                    b+c+d & a+b+d
                                                           \end{pmatrix}\] using the usual Hamming weight $w_{\tt Ham}$ on $\FF_4$ extended component-wise, and the Bachoc weight $w_{\tt B}$ on $M_2(\FF_2)$ such that $w_{\tt Ham}(\alpha)=w_{\tt B}(\phi(\alpha))$ for all $\alpha$ in $\FF_4^2$. Here $\omega$ is a root of the monic irreducible polynomial $x^2+x+1 \in \FF_2[x]$ such that $\FF_4$ is seen as an extension of $\FF_2$ by $\omega$.  The Bachoc weight on $M_2(\FF_p)$ as given in \cite{bac} is defined as follows,
\begin{gather*}
		   w_{\tt B}(A) = \begin{cases}
       		 0 & {\rm if} \ A = {\bf 0} \\
                         1 & {\rm if} \ A\in GL(2,p)\\
     		 p & \  otherwise.\
      		  \end{cases}
		\end{gather*}
The study of codes over $\ZZ_4$ and $M_2(\FF_2)$ reveals the importance of weight functions that are different from the Hamming weight. Here we derive the homogeneous weight on $M_2(\FF_p)$ using the formula introduced by T. Honold for arbitrary finite Frobenius rings \cite{hon}. Likewise we extend the definition of the Lee weight on $\FF_2+u\FF_2$, $u^2=0$ given in \cite{bon:uda} to the finite chain ring $\FF_{p^2}+u\FF_{p^2}$, $u^2=0$. The connection between the minimal left ideals and the idempotent elements of $M_2(\FF_p)$ is used to generalize the homogeneous weight on $M_2(\FF_p)$. We also employ the well known representation of the field by matrices by giving a unital embedding $\tau$ from $\FF_{p^2}$ to $M_2(\FF_p)$ to construct a non-commutative ring that is isomorphic to $M_2(\FF_p)$ and is a left $\FF_{p^2}$-vector space.  This ring is denoted by $\mathcal{F}_{p^2} + \mathbf{v}_p\mathcal{F}_{p^2}$ where $\mathbf{v}_p$ is an involution in $M_2(\FF_p)$ and the elements of $\mathcal{F}_{p^2}$ come from $M_2(\FF_p)$ such that $\tau(\FF_{p^2})\cong \mathcal{F}_{p^2}$.  The unital embedding $\tau$ comes from a characterization of $\FF_{p}$ in terms of an irreducible polynomial $f(x)=x^2+x+(p-1) \in \FF_p[x]$.  The property if this polynomial restricts our study to the case where $p\equiv$ $2$ or $3$ mod $5$. As a consequence certain structural properties of cyclic codes over $M_2(\FF_p)$ that are similar to those of cyclic codes over $M_2(\FF_2)$ are derived. The structure theorems used the transformation of the non-commutative ring $\mathcal{F}_{p^2}+\mathbf{v}_p\mathcal{F}_{p^2}$ to $\mathcal{F}_{p^2}+\mathbf{u}_p\mathcal{F}_{p^2}$ by introducing a matrix $\mathbf{i}_p \in M_2(\FF_p)$ such that $\mathbf{u}_p=\mathbf{i}_p+\mathbf{v}_p$, where $\mathbf{u}_p^2$ is the zero matrix.  Also we define a left $\FF_p$-module isometry from $M_2(\FF_p)$ to $\FF_{p^2}+u\FF_{p^2}$ using their respective Bachoc weight and Lee weight. 

\section{Homogeneous Weight on $M_2(\FF_p)$}
Let $R$ be a finite ring and $\RR$ the set of real numbers. A weight function $w\colon R \longrightarrow \RR$ is called {\it left homogeneous} provided $w(0)=0$ and the following hold:\vskip .1in
\begin{enumerate}
	\item[(H1)] If $Rx=Ry$ for $x,y \in R$, then $w(x)=w(y).$
	\item[(H2)] There exists $\Gamma > 0$ such that for every nonzero $x\in R$ there holds
	\[\sum_{y\in Rx} w(y) = \Gamma |Rx|.\]
\end{enumerate}
\bigskip
The definition for a right homogeneous weight follows analogously, and we say that $w$ is {\it homogeneous} if it is both left homogeneous and right homogeneous. The number $\Gamma$ is called the {\it average value} of $w$. The weight $w$ is said to be {\it normalized} if $\Gamma=1$. It is well known that the normalized homogeneous weight on $\FF_{q}$, $q=p^r$, is given by 
\begin{gather*}
		   w_{\tt nhom}(x) = \begin{cases}
       		 0 & {\rm if} \ x = 0 \\
     		 \dfrac {q}{q-1} & {\rm if} \ x \ne 0. \\
      		  \end{cases}
		\label{nhom}
	\end{gather*}
This idea comes from the generalization of the homogeneous weight on a finite chain ring \cite{gre:sch}. But our goal is to give a generalization of the homogeneous weight on $M_2(\FF_p)$ which is not a finite chain ring but is a finite (non-commutative) Frobenius ring. We shall use the generating character instead of the M\"obius inversion formula for homogeneous weight that was employed in \cite{gre:sc2}.

For a finite Frobenius ring $R$, Honold \cite{hon} observed that every homogeneous weight on $R$ with generating character $\chi$ must have the form \[w: R\ \longrightarrow \RR, \quad x\mapsto \Gamma \Biggl[1-\dfrac{1}{|R^{\times}|}\sum_{u\in R^{\times}} \chi(ux)\Biggl]\]
where $R^{\times}$ is the group of units of $R$. Note that every finite Frobenius ring has a generating character \cite{woo}. The generating character of $M_n(\FF_q)$ is  
\[ \chi(A)=exp\Biggl\{\dfrac{2\pi i\cdot tr(Tr(A))}{p}\Biggl\}\] where $tr$ is the trace map from $\FF_q$ down to $\FF_p$, that is, $tr(\alpha)=\alpha+\alpha^p+\dots+\alpha^{p^{r-1}}$ for $\alpha\in \FF_q$, and $Tr$ is the classical trace of the matrix $A \in M_n(\FF_q)$.
The homogeneous weight on $M_n(\FF_{q})$  is given by 
\[w: M_n(\FF_{q}) \longrightarrow \RR, \quad A\mapsto \Gamma \Biggl[1-\dfrac{1}{|GL(n,q)|}\sum_{u\in GL(n,q)} \chi(uA)\Biggl]\]
where $GL(n,q)$ is the group of nonsingular matrices in $M_n(\FF_{q})$. It is known that $|GL(n,q)|=q^{n(n-1)/2}\prod_{i=1}^n (q^i-1)$ \cite{big:whi}.

The main concern in this section is to derive the homogeneous weight on $M_2(\FF_{p})$. First we discuss the structure of $M_2(\FF_{p})$.

\begin{remark} The matrix ring $M_n(\FF_{q})$ has no proper ideals but it has proper left ideals \cite{hun}.  In particular $M_2(\FF_{p})$ has $p+1$ minimal left ideals \cite{bac}.  This is essential in this section so we take it as a theorem. 
\end{remark}

\begin{theorem} $M_2(\FF_{p})$ has $p+1$ minimal left ideals and each minimal left ideal contains $p^2$ elements.
\end{theorem}
\pf
Let $A \in M_2(\FF_{p})$ where $A={\begin{pmatrix}
                                                      a_0 & a_1  \\
                                                      a_2 & a_3 
                                           \end{pmatrix}}$. Note that $ {\begin{pmatrix}
                                                      1 & r  \\
                                                      0 & 0 
                                           \end{pmatrix}}$ and $ {\begin{pmatrix}
                                                      0 & 0  \\
                                                      0 & 1 
                                           \end{pmatrix}}$ are nonzero nonunit idempotents of $M_2(\FF_{p})$ where $r \in \FF_{p}$.  Thus the proper left ideals are of the form $ {\begin{pmatrix}
                                                     a_ 0 & ra_0  \\
                                                      a_2 & ra_2
                                           \end{pmatrix}}$ and $ {\begin{pmatrix}
                                                      0 & a_1  \\
                                                      0 & a_3 
                                           \end{pmatrix}}.$ Hence there are $p+1$ minimal left ideals in $M_2(\FF_{p})$ since the intersection of any two minimal left ideals of $M_2(\FF_{p})$ is the zero matrix and every minimal left ideal of $M_2(\FF_{p})$ has $p^2$ elements.
\Q\Box

In order to generalize the homogeneous weight on $M_2(\FF_p)$ we need to get the value of the sum $ \sum_{u\in GL(2,p)}\chi(uA)$ where $A\in M_2(\FF_{p})$.  The case when $A$ is the zero matrix is obvious.  Theorem~\ref{theorem:invertible} below deals with the invertible matrices while Theorem~\ref{theorem:zerodivisors} involves the zero divisors.  

\begin{theorem}\label{theorem:invertible} $\sum_{u \in GL(2,p)} \chi(u)=\sum_{u \in GL(2,p)} \chi(uA)= p$ where $A \in GL(2,p)$.
\end{theorem}

\pf
Let $D$ be the set of all the zero divisors in $M_2(\FF_{p})$. We have $\sum_{A \in M_2(\FF_{p})} \chi(A) =0$ \cite {hon}.
 So, 
\[
\sum_{u \in GL(2,p)} \chi(u) = -\sum_{B \in D} \chi(B) - \chi(0) 
\]\nonumber
and since $M_2(\FF_{p})$ has $p+1$ minimal left ideals, $\chi_{I_L}(A)=\chi(A)$ for all $A\in {I_L}$ and $\chi_{I_L}(0)=1$ in \cite{hon}, where $\chi_{I_L}$ is a character of the minimal left ideal $I_L$ of $M_2(\FF_{p})$. Hence,
\begin{equation}\nonumber
\sum_{u \in GL(2,p)} \chi(u)=-(p+1) \sum_{B \in I_L \backslash \{0\}} \chi_{I_L}(B) -1  =-(p+1)(-1)-1 =p. 
\end{equation} \Q\Box
\begin{theorem}\label{theorem:zerodivisors} $\sum_{u_k \in GL(2,p)} \chi(u_kB)=p-p^2$ for all $B\in I_L \backslash \{0\}$.
\end{theorem}
\vskip.1in
\pf

\[\begin{aligned}
-\sum_{u_k \in GL(2,p)} \chi(u_kB)
&= \Biggl [\sum_{B_j \in I_L\backslash \{0\}} \chi(B_j)\Biggl] \Biggl[\sum_{u_k \in GL(2,p)} \chi (u_kB) \Biggl]\\
&=\sum_{B_j \in I_L \backslash \{0\}} \sum_{u_k \in GL(2,p)} \chi (u_kB)\chi(B_j)\\
&=\sum_{B_j \in I_L \backslash \{0\}} \sum_{u_k \in GL(2,p)} \chi (u_kB + B_j)\\
&= \sum_{u_k \in GL(2,p)} \sum_{B_j \in I_L \backslash \{0\}}\chi (u_kB + B_j)
\end{aligned}\]
For each $u_r\in GL(2,p)$, there exists $B_s \in I_L\backslash \{0\}$ such that $u_rB+B_s = 0$  (Note: $B_s$ is not unique for every $u_r$). Thus, given that $u_kB+B_j \ne 0 $ we have 

\[\begin{aligned}
-\sum_{u_k \in GL(2,p)} \chi(u_kB)
&=\sum_{u_r \in GL(2,p)} \chi (u_rB + B_s) + \sum_{u_k \in GL(2,p)} \sum_{B_j \in I_L \backslash \{0\}}\chi (u_kB + B_j) \\
&=\sum_{u_r \in GL(2,p)} \chi (0) + \sum_{u_k \in GL(2,p)} \sum_{B_j \in I_L \backslash \{0\}}\chi (u_kB + B_j)\\
&=|GL(2,p)| + \sum_{u_k \in GL(2,p)} \sum_{B_j \in I_L \backslash \{0\}}\chi (u_kB + B_j).
\end{aligned}\]
\noindent
For every $B_t \in I_L\backslash \{ 0, B_s\}$ we have $u_rB+B_t$ $\in I_L \backslash \{ 0, u_rB\}$  and for fixed $B_t$ and $u_r$ we can always find $l$ such that $u_l\ne u_r$ and $u_rB + B_t= u_rB.$  Thus, we can collect all the elements of $I_L\backslash \{ 0\}$.  And since $|I_L \backslash \{0\}|$ divides $|GL(2,p)|$,

\[\begin{aligned}
-\sum_{u_k \in GL(2,p)} \chi(u_kB)
&= |GL(2,p)| + \dfrac {|GL(2,p)||I_L\backslash \{0\}| - |GL(2,p)|}{|I_L \backslash \{0\}|} \sum_{B_j \in I_L \backslash \{ 0\}} \chi (B_j) \\
&=(p^2-p)(p^2-1)+(p^2-p)(p^2-2)(-1)\\
&= p^2-p.
\end{aligned}\]
Thus,  $\sum_{u_k \in GL(2,p)} \chi(u_kB)=p-p^2.$\Q\Box

\begin{theorem} The homogeneous weight on $M_2(\FF_{p})$ is given by
	 \begin{gather*}
		   w_{\tt hom}(A) = \begin{cases}
       		 0 & {\rm if} \ A = {\bf 0} \\
     		\Gamma\Biggl( 1-  \dfrac {1}{(p^2-1)(p-1)}\Bigg)& {\rm if} \ A \in GL(2,p) \\
      		 \Gamma\Biggl( \dfrac {p^2}{p^2-1}\Biggl) &  \ otherwise .\\
      		 \end{cases}
		\label{hom}
	\end{gather*}
\end{theorem}
\pf
The proof is straightforward from the two preceding theorems. 
\Q\Box
\begin{table}\caption{Bachoc Weight and Normalized Homogeneous Weight on $M_2(\FF_2)$}
\begin{center}
\begin{tabular}{|c|c|c||c|c|c|}
\hline $M_2(\FF_p)$  & $w_{\tt B}$ & $w_{\tt nhom}$ & $M_2(\FF_p)$  & $w_{\tt B}$ & $w_{\tt nhom}$\\
\hline $\begin{pmatrix} 0&0\\ 0&0 \end{pmatrix}$  & $0$ & $0$ & $\begin{pmatrix} 0&1\\ 0&1 \end{pmatrix}$ & $2$ & $4/3$\\
$\begin{pmatrix} 1&0\\0&1 \end{pmatrix}$ & $1$ & $2/3$ & $\begin{pmatrix} 1&1\\ 0&0 \end{pmatrix}$ & $2$ & $4/3$\\
$\begin{pmatrix} 0&1 \\ 1&1 \end{pmatrix}$ & $1$ &$2/3$ & $\begin{pmatrix} 0&0\\ 0&1 \end{pmatrix}$ & $2$ & $4/3$\\
$\begin{pmatrix} 1&1 \\ 1&0 \end{pmatrix}$ & $1$ &$2/3$ & $\begin{pmatrix} 1&0\\ 0&0 \end{pmatrix}$ & $2$ & $4/3$\\
$\begin{pmatrix} 1&0 \\ 1&1 \end{pmatrix}$ & $1$ &$2/3$ & $\begin{pmatrix} 1&1\\ 1&1 \end{pmatrix}$ & $2$ & $4/3$\\
$\begin{pmatrix} 1&1 \\ 0&1 \end{pmatrix}$ & $1$ &$2/3$ & $\begin{pmatrix} 0&0\\ 1&1 \end{pmatrix}$ & $2$ & $4/3$\\
$\begin{pmatrix} 0&1 \\ 1&0 \end{pmatrix}$ & $1$ &$2/3$ & $\begin{pmatrix} 1&0\\ 1&0 \end{pmatrix}$ & $2$ & $4/3$\\
$\begin{pmatrix} 0&0 \\ 1&0 \end{pmatrix}$ & $2$ &$4/3$ & $\begin{pmatrix} 0&1\\ 0&0 \end{pmatrix}$ & $2$ & $4/3$\\
\hline
\end{tabular}
\end{center}
\end{table}
%\vskip.1in
\section{Lee Weight on $\FFF_{p^2}+u\FFF_{p^2}$, $u^2=0$}

In \cite{bon:uda} the Lee weight $w_{\tt L}$ of $x=(x_1,\dots,x_n) \in (\FF_2+u\FF_2)^n$ is defined as $n_1(x)+2n_2(x)$, where $n_2(x)$ and $n_1(x)$ are, respectively, the number of $u$ symbols and the number of $1$ or $1+u$ symbols present in $x$.  So when $n=1$, $w_{\tt L}(0)=0$, $w_{\tt L}(1)=w_{\tt L}(1+u)=1$ and $w_{\tt L}(u)=2$.

Consider the finite chain ring $\FF_3+u\FF_3$, $u^2=0$ then we can define $w_{\tt L}(x)=n_1(x)+3n_2(x)$, for all $x\in \FF_3+u\FF_3$, where $n_2(x)$ and $n_1(x)$ are, respectively, the number of $u$ symbols and the number of $1$ or $1+u$ symbols present in $x$, as can be seen in Table 
\ref{lee weight F3+uF3}.
\begin{table}\caption{Lee weight on $\FF_3+u\FF_3$, $u^2=0$}
\label{lee weight F3+uF3}
\begin{center}
\begin{tabular}{|c|c|}
\hline $\FF_3+u\FF_3$ & $w_{\tt L}$ \\
\hline $0$ & $0$\\
$1$ & $1$\\
$2$ & $1$\\
$1 + u$ & $1$\\ 
$2+2u=2(1+u)$ & $1$\\
$u$ & $3$\\
$2+u$&$3$ \\
$2u$&$3$ \\
$1+2u$ & $3$\\
\hline
\end{tabular}
\end{center}
\end{table}
\vskip.1in
Now consider the subset $\mathcal{B}_2$ of $\FF_4+u\FF_4$, $u^2=0$ where \[\mathcal{B}_2=\{(\alpha a_1 + \alpha b_1 \omega)+u(\beta a_1 +\beta b_1 \omega)| \alpha=1, a_1,b_1,\beta\in \FF_2\}.\] Similarly we can define the Lee weight on $\FF_4+u\FF_4$, $u^2=0$ to be $w_{\tt L}(x)=n_1(x)+2n_2(x)$ where again $n_2(x)$ and $n_1(x)$ are, respectively, the number of $u$ symbols and the number of $1$ or $1+u$ symbols present in $x$. 
\begin{gather*}
		   w_{\tt L}(x) = \begin{cases}
       		 0 & {\rm if} \ x = {\bf 0} \\
                         1 & {\rm if} \ A\in \mathcal{B}_2\backslash \{0\}\\
     		 2 & \  otherwise\
      		  \end{cases}
		\end{gather*}
This can also be seen in Table \ref{Lee weights on F4+uF4}. In general we can define the Lee weight on $\FF_{p^2}+u\FF_{p^2}$, $u^2=0$ as $w_{\tt L}(x)=n_1(x)+pn_2(x).$
\begin{gather*}
		   w_{\tt L}(x) = \begin{cases}
       		 0 & {\rm if} \ x = {\bf 0} \\
                         1 & {\rm if} \ A\in \mathcal{B}_p\backslash \{0\}\\
     		 p & \  otherwise\
      		  \end{cases}
		\end{gather*}
\noindent
where $\mathcal{B}_p=\{(\alpha a_1 + \alpha b_1 \omega)+u(\beta a_1 +\beta b_1 \omega)| \alpha\in \FF_p^\times, a_1,b_1,\beta\in \FF_p\}.$

\begin{table}\caption{Lee Weight on $\FF_4+u\FF_4$, $u^2=0$}
\label{Lee weights on F4+uF4}
\begin{center}
\begin{tabular}{|c|c|}
\hline $\FF_4+u\FF_4$ & $w_{\tt L}$ \\
\hline $0$ & $0$\\
$1$ & $1$\\
$\omega$ & $1$\\
$1 + \omega$ & $1$\\ 
$1+u$ & $1$\\
$\omega+u\omega=\omega(1+u)$ & $1$\\
$(1+\omega)+u(1+\omega)=(1+\omega)(1+u)$&$1$ \\
$u$&$2$ \\
$\omega + u$ & $2$\\
$(1+\omega)+ u$ & $2$\\
$u\omega$ & $2$ \\
$1+u\omega$ & $2$\\
$(1+\omega)+u\omega$ & $2$\\
$u(1+\omega)$ & $2$ \\
$1+u(1+\omega)$ & $2$ \\
$\omega+u(1+\omega)$ & $2$ \\\hline
\end{tabular}
\end{center}
\end{table}
\section{$\FFF_{p^2}$-Linear Map}
In this section we give the conditions on the finite field $\FF_{p}$ for the polynomial $f(x)=x^2+x+(p-1)$ to be irreducible over $\FF_{p}$. Using the well known representation of fields by matrices, Theorem \ref{representation of field by matrices} shows the corresponding cyclic algebra that is isomorphic to $M_2(\FF_{p})$ and is a left $\FF_{p^2}$-vector space. 
\begin{lemma}\label{lemma:irreducible}
Let $p \equiv$ $2$ or $3$ (mod $5$) then the polynomial $f(x)=x^2+x+(p-1)$ is irreducible over $\FF_{p}$.
\end{lemma}
\pf
The case when $p=2$ is trivial. Note that the discriminant of the polynomial $f(x)$ is equal to  $5 \in \FF_{p}$. Then $f(x)$ is reducible over $\FF_{p}$ if  there exists $y \in \FF_{p}$ such that  $y^2 \equiv$ $5$ (mod $p$).  By the Law of Quadratic Reciprocity of elementary number theory, when $p$ is odd, $y^2 \equiv$ $5$ (mod $p$) is solvable if and only if $p \equiv$ 1 or -1 (mod $5$).
\Q\Box

\begin{theorem}\label{representation of field by matrices}Let $f(x)=\sum_{i=0}^n a_ix^i \in \FF_q[x]$ be a monic irreducible polynomial.  Then the mapping $\pi\colon \FF_q[x] \rightarrow M_n(\FF_q)$, $g(x) \mapsto g(X)$ induces a unital embedding of $\FF_q[x]/(f)$ into $M_n(\FF_q)$ where 
\[X=\begin{pmatrix}
 0&0&\cdots&0& -a_0\\
1&0&\cdots&0&-a_1\\
0&1&\cdots&0&-a_2\\
\vdots&\ddots&\ddots&\vdots&\vdots\\
0&0&\cdots&1&-a_{n-1}
\end{pmatrix}.\]
\end{theorem}
\begin{remark} The matrix $X$ is known as the {\it companion matrix}.
\end{remark}
\begin{corollary} Let $\FF_{p^2}=\FF_{p}[\omega]$ where $\omega^2+\omega+(p-1)=0$ then $\tau\colon \FF_{p^2} \longrightarrow M_2(\FF_{p})$ defined by

\[a+b\omega \mapsto \begin{pmatrix}
				a & b\\
				b& a+(p-1)b\end{pmatrix}\] is an embedding.
\end{corollary}
\pf
The proof follows immediately from Lemma~\ref{lemma:irreducible} and Theorem~\ref{representation of field by matrices}.
\Q\Box
\begin{theorem}
If $\omega$ is a root of $f(x)=x^2+x+(p-1)$ then $\omega^p\equiv (p-1)\omega+(p-1) (\bmod (\omega^2+\omega+(p-1))$.
\end{theorem}
\pf
First we show that $(p-1)\omega+(p-1)$ is also a root of $f(x)$, that is, 
\[\begin{aligned}
f[(p-1)\omega+(p-1)]
&= [(p-1)\omega+(p-1)]^2+[(p-1)\omega+(p-1)]+(p-1)\\
  &=[(p-1)^2\omega^2+2\omega+1]+[(p-1)\omega+(p-1)]+(p-1)\\
  &=\omega^2+2\omega+1-\omega-2\\
  &=\omega^2+\omega+(p-1)\\
  &=0.
\end{aligned}\]

Now, let $h(x)=x^p$. By the Division Algorithm, there exist $g(x)$ and $r_1x+r_2$ such that $h(x)=g(x)f(x)+r_1x+r_2$ where $r_1x+r_2$ is the remainder when $h(x)$ is divided by $f(x)$.  Since $\omega$ and $(p-1)\omega+(p-1)$ are roots of $f(x)$ then we have $$\omega^p=r_1\omega+r_2$$ and $$ [(p-1)\omega+(p-1)]^p=r_1[(p-1)\omega+(p-1)]+r_2 $$ or equivalently, $$ (p-1)\omega^p+(p-1)=r_1(p-1)\omega+r_1(p-1)+r_2.$$ Since the characteristic of $\FF_p$ is $p$, then
\[[(p-1)\omega+(p-1)]^p=[(p-1)\omega]^p+(p-1)^p=[(p-1)^p\omega^p]+(p-1)^p.\] By Fermat's Little Theorem,
\[[(p-1)^p\omega^p]+(p-1)^p=(p-1)\omega^p+(p-1).\]
Adding equations $\omega^p=r_1\omega+r_2$ and $(p-1)\omega^p+(p-1)=r_1(p-1)\omega+r_1(p-1)+r_2$ modulo $p$, the resulting equation is $(p-1)=r_1(p-1)+2r_2$ or simply $r_1+(p-2)r_2=1$.
\vskip.1in
Note that $gcd(1,p-2)=1$.  And we have $1=(p-1)-(p-2)=(p-1)+(p-2)(p-1)$.  So, $r_1=p-1$ and $r_2=p-1$.  Thus, $\omega^p\equiv (p-1)\omega+(p-1) (\bmod (\omega^2+\omega+(p-1))$.
\Q\Box

\vskip.1in
\begin{theorem}
$\tau^p(\omega)=\begin{pmatrix} p-1&p-1\\
                                                         p-1&0
                                \end{pmatrix}.$
\end{theorem}
\pf
Since $\tau$ is a homomorphism we have $\tau(\omega^p)=\tau^p(\omega)$. And from Corollary 4.4 and Theorem 4.5 we have 
\[\begin{aligned}\tau^p(\omega)
      &=\tau[(p-1)\omega+(p-1)]\\
      &=\tau[(p-1)\omega]+\tau(p-1)\\
      &=\tau(p-1)\tau(\omega)+\tau(p-1)\\
      &=\begin{pmatrix} p-1&0\\
                                      0&p-1 \end{pmatrix}
           \begin{pmatrix} 0&1\\
                                      1&p-1 \end{pmatrix}
         +\begin{pmatrix} p-1&0\\
                                      0&p-1 \end{pmatrix}\\
      &=\begin{pmatrix} 0&p-1\\
                                     p-1&1 \end{pmatrix}
           +\begin{pmatrix} p-1&0\\
                                        0&p-1 \end{pmatrix}\\
       &=\begin{pmatrix} p-1&p-1\\
                                      p-1& 0 \end{pmatrix}.
\end{aligned}\]
\Q\Box

\begin{theorem} Let $\mathcal{F}_p$ be the set of all scalar matrices in $M_2(\FF_p)$, $p\equiv2$ or $3\bmod 5$, $\tau(\FF_{p^2})=\mathcal{F}_{p^2}$ and $\mathbf{v}_p=\begin{pmatrix} 1&0\\ p-1&p-1 \end{pmatrix}$. Then 
$\mathbf{v}_p\tau(\omega)=\tau^p(\omega)\mathbf{v}_p$, $\mathcal{F}_p[\tau(\omega)]=\mathcal{F}_{p^2}$ and 
$M_2(\FF_p)=\mathcal{F}_{p^2}+\mathbf{v}_p\mathcal{F}_{p^2}.$ 
\end{theorem}
\pf It is easy to show that $\mathbf{v}_p\tau(\omega)=\tau^p(\omega)\mathbf{v}_p$ and $\mathcal{F}_p[\tau(\omega)]=\mathcal{F}_{p^2}$ since $\tau^2(\omega)+\tau(\omega)+\tau(p-1)=\begin{pmatrix} 0&0\\ 0&0\end{pmatrix}.$ 
\[\begin{aligned}M_2(\FF_p)&=\mathcal{F}_{p^2}+\mathbf{v}_p\mathcal{F}_{p^2}\\
&=\Biggl\{\begin{pmatrix} a+c&b+d \\
b-c-d&a-b-c)\end{pmatrix}| a,b,c,d \in \FF_p\Biggl\}.
\end{aligned}\]
\Q\Box

\section{Cyclic Codes over $M_2(\FF_p)$}
Structure theorems for cyclic codes over $\mathcal{A}_2=M_2(\FF_2)$ were established in \cite{ala:sbo:sol:yem} by introducing two matrices $\tau(\omega)$ and ${\tt v}$ in $\mathcal{A}_2$ satisfying the relation ${\tt v}\tau(\omega)=\tau^2(\omega){\tt v}$.  A possible choice would be those given by Bachoc~\cite{bac} which are 
${\tt v}= \begin{pmatrix} 0&1 \\
			                1&0
		  \end{pmatrix}$ 
and $\tau(\omega)=\begin{pmatrix} 0& 1\\
				  1&1	
		\end{pmatrix}$
\noindent
such that $\mathcal{A}_2=\mathcal{F}_2[\tau(\omega)]+ {\tt v} \mathcal{F}_2[\tau(\omega)]$ where $\mathcal{F}_2[\tau(\omega)]=\mathcal{F}_4\cong \FF_4$ and ${\tt v} \ne \mathbf{v}_2$.  Setting ${\tt u}= \tau(1)+{\tt v}$ gives ${\tt u}^2= \begin{pmatrix} 0&0\\
					0&0 \end{pmatrix}$ and $A_2=\mathcal{F}_4 + {\tt u} \mathcal{F}_4$.  Alamadhi et.al. \cite{ala:sbo:sol:yem} used the ring $\mathcal{F}_4 + {\tt u} \mathcal{F}_4$ to develop structure theorems for cyclic codes over $M_2(\FF_2)$ by simply extending from cyclic codes over $\FF_2 +u\FF_2$, $u^2=0$ \cite{bon:uda}. 
					
It seems that a construction of cyclic codes over $\FF_p+u\FF_p$, $u^2=0$, will result in the construction of cyclic codes over $\mathcal{A}_p=M_2(\FF_p)$. Fortunately, Qian, Zhang and Zhu \cite{qia:zha:zhu} solved an open-ended question given in \cite{bon:uda}, that is, to extend the cyclic codes over $\FF_2+u\FF_2$, $u^2=0$ to $\FF_p+u\FF_p +\dots +u^{k-1}\FF_p$, $u^{k}=0$.  Thus, the case when $k=2$ gives the cyclic codes over $\FF_p+u\FF_p$, $u^2=0$. 

\begin{figure}[ht]

	{\centering

	$\FF_{p^2}+u\FF_{p^2}$\\ 
	$\vert$ \\
	$u\FF_{p^2}$\\
	$\lvert$ \\
	$(0)$\\
            
	\caption {Lattice of ideals of $\FF_{p^2}+u\FF_{p^2}, u^2=0$}
                                                              
  }
	
\end{figure}

\noindent Let $p\equiv 2$ or $3$ (mod $5$), $\mathbf{i}_p= \begin{pmatrix} p-1&0\\
				0&1
		\end{pmatrix}$ and $\mathbf{u}_p=\mathbf{v}_p+\mathbf{i}_p$. Then $\mathbf{u}_p^2=\begin{pmatrix} 0&0\\
                                                      0&0 \end{pmatrix}$ and
\[\begin{aligned}
\mathcal{A}_p&=\mathcal{F}_{p^2}+\mathbf{u}_p\mathcal{F}_{p^2}\\
&=\Biggl\{\begin{pmatrix} a&b\\b-c&a-b-d \end{pmatrix}|a,b,c,d \in \FF_p\Biggl\}.
\end{aligned}\]
\vskip.1in
\begin{figure}[htb]
	\centering

	$\mathcal{F}_{p^2}+ \mathbf{u}_p\mathcal{F}_{p^2}$\\ 
	$\lvert$ \\
	$(0)$
	\caption{Lattice of ideals of $\mathcal{F}_{p^2}+\mathbf{u}_p\mathcal{F}_{p^2}$,$\mathbf{u}_p^2$ is the zero matrix}
	
\end{figure}

Let $\mathcal{A}_p[X]$ be the ring of polynomials over $\mathcal{A}_p$.  We have a natural homomorphic mapping from $\mathcal{A}_p$ to its field $\mathcal{F}_{p^2}$.  For any $a \in \mathcal{A}_p$, let $\hat{a}$ denote the polynomial reduction modulo $\mathbf{u}_p$.  Now define a polynomial reduction mapping $\mu\colon \mathcal{A}_p[X] \longrightarrow \mathcal{F}_{p^2}[X]$ such that
\[f(X)=\sum_{i=0}^r a_iX^j \mapsto \sum_{i=0}^r \hat{a_i}X^j.\]
\indent
A monic polynomial $f$ over $\mathcal{A}_p[X]$ is said to be a basic irreducible polynomial if its projection $\mu(f)$ is irreducible over $\mathcal{F}_{p^2}[X]$. An $\mathcal{A}_p$-linear code $C$ of length $n$ is an $\mathcal{A}_p$-submodule of $\mathcal{A}_p^n$.  As left modules we have the expansion $\mathcal{R}_{p,n}=\mathcal{A}_p[x]/(x^n-1)= \oplus_{j=1}^t \mathcal{A}_{p,j}$, where the $\mathcal{A}_{p,j}=\mathcal{A}_p[x]/(f_j)$ are quotient $\mathcal{A}_p$-modules and $x^n-1=\prod_{j=1}^t f_j$ where $f_j$'s are irreducible polynomials over $\mathcal{F}_{p^2}.$

We shall prove the lemma and the theorem below using the same techniques in \cite{ala:sbo:sol:yem} and \cite{qia:zha:zhu} given the condition that $p$ is not divisible by $n$.
\begin{lemma}
If $f$ is an irreducible polynomial over $\mathcal{F}_{p^2}$ the only left $\mathcal{A}$-modules of $\mathcal{R}_p(f)=\mathcal{A}_p[X]/(f)$ are $(\tau(0))$, $(\mathbf{u}_p)$ and $(\tau(1))$.  In particular this quotient ring is a non-commutative chain ring.
\end{lemma}
\pf
Let $I\ne (\tau(0))$ be an ideal of $\mathcal{R}_p(f)$.  Pick $g$ in $\mathcal{A}_p[X]$ such that $g+(f) \in I$, but $g\notin (f)$.  Because $f$ is irreducible the $gcd$ of $\mu g$ and $f$ can only take two values, $\tau(1)$ and $f$.  In the first case $g$ is invertible mod $f$ and $I=(\tau(1))=\mathcal{R}_p(f)$.  If this does not happen, $I\subseteq \mathbf{u}_p+(f)$.  To show the reverse inclusion, let $g=\mathbf{u}_pr$ with $\mathbf{u}_pr+(f) \subseteq I$ and $\mathbf{u}_pr+(f) \ne \tau(0)$.  We can assume by the latter condition that $\mu r \notin (f)$.  Hence by the irreducibility of $f$ we have that $gcd(\mu r, f)=\tau(1)$.  This entails the existence of $a,b,c \in \mathcal{A}_p[X]$ such that $ra+fb=\tau(1)+\mathbf{u}_pc.$ Multiplying both sides by $\mathbf{u}_p$ we get $\mathbf{u}_pra=\mathbf{u}_p+\mathbf{u}_pfb.$ The left hand side is in $I$, a right sided ideal. Thus the reverse inclusion follows.
\Q\Box
\vskip.1in
\begin{theorem}
Suppose $C$ is a cyclic code of length $n$ over $\mathcal{A}_p=\mathcal{F}_{p^2}+\mathbf{u}_p\mathcal{F}_{p^2}$ where $p$ is not divisible by $n$. Then there are unique monic polynomials $F_0, F_1, F_2$ such that $C=\langle \hat{F}_1,\mathbf{u}_p\hat{F}_2\rangle$, where $F_0F_1F_2=X^n-\begin{pmatrix} 1&0\\0&1 \end{pmatrix}$, $\hat{F}_1=F_0F_2$, $\hat{F}_2=F_0F_1$, and $\lvert C \rvert=p^{2s}$ where $s=2 \deg F_1+\deg F_2.$
\end{theorem}
\pf
Let $X^n-\begin{pmatrix} 1&0\\0&1 \end{pmatrix}=f_1f_2\dots f_r$  be the unique factorization of $X^n-\begin{pmatrix} 1&0\\ 0&1 \end{pmatrix}$ into a product of monic basic irreducible pairwise coprime polynomials. Note that $C$ is a direct sum of right $\mathcal{A}_p$-modules of the form $(\mathbf{u}_p^j\hat{f}_i)$, $0\le j \le 1$, $0\le i\le r$ where $\hat{f}_i=\prod_{j=1, j\ne i}^nf_j$.  After reordering, we can assume that $C$ is a direct sum of any of the following
\[(\hat{f}_{t_1+1}), (\hat{f}_{t_1+2}), \dots , (\hat{f}_{t_1+t_2}), (\mathbf{u}_p\hat{f}_{t_1+t_2+1}), \dots, (\mathbf{u}_p\hat{f}_r).\] That is,
\[C=\langle f_1f_2f_3\dots f_{t_1}f_{t_1+t_2+1}\dots f_r, \mathbf{u}_pf_1f_2f_3 
\dots f_{t_1+t_2}f_{t_1+t_2+t_3}\rangle. \]
Let $\hat{F}_1= f_1f_2f_3\dots f_{t_1}f_{t_1+t_2+1}\dots f_r$  and 
$\hat{F}_2=f_1f_2f_3 \dots f_{t_1+t_2}f_{t_1+t_2+t_3}.$
where $t_1,t_2 \ge 0$ and $t_1+t_2+1\le r$.
\vskip.1in
\noindent
Then
\[ F_i = \begin{cases} 1 & t_{i+1}=0\\
                                    f_{t_0+t_1+\dots +t_i+1}\dots f_{t_0+t_1+\dots + t_{i+1}} & t_{i+1}\ne 0,
\end{cases} \]
where $t_0=0$ , $0\le i \le 2.$
\vskip.1in
\noindent
Then by our construction, it is clear that $C=\langle \hat{F}_1, \mathbf{u}_p\hat{F}_2 \rangle$ and $X^n-\begin{pmatrix} 1&0\\ 0&1 \end{pmatrix}=F_0F_1F_2=f_1f_2\dots f_r.$
\vskip.1in
To prove uniqueness, we assume that $G_0,G_1,G_2$ are pairwise coprime monic polynomials in $\mathcal{A}_p[X]$ such that $G_0G_1G_2=X^n-\begin{pmatrix} 1&0\\ 0&1 \end{pmatrix}$ and $C=\langle \hat{G}_1, \mathbf{u}_p\hat{G}_2 \rangle$.  Thus, $C= (\hat{G}_1)+(\mathbf{u}_p\hat{G}_2)$.  Now there exist nonnegative integers $m_0=0,m_1,\dots, m_{d+1}$ with $m_0+m_1+\dots+m_{d+1}=r$, and a permutation of $\{f_1,f_2, \dots, f_r \}$ such that $G_i=f_{m_0+\dots + m_{i}+1}\dots f_{m_0+\dots+m_{i+1}}$ for $i=0,1,2.$  Hence,
\[C=(\hat{f}_{m_1+1})\oplus \dots \oplus (\hat{f}_{m_1+m_2})\oplus (\mathbf{u}_p\hat{f}_{m_1+m_2+1}) \dots \oplus (\mathbf{u}_p\hat{f}_r) .\]
It follows that $m_i=t_i$ for $i=0,1,2.$  Furthermore, $(f_{m_0+\dots+m_d+1}, \dots , f_{m_0+\dots + m_{d+1}})$ is a permutation of $\{f_{t_0+\dots+t_d+1}, \dots, f_{t_0+\dots + t_{d+1}}\}.$  Therefore, $F_i=G_i$ for $i=0,1,2.$  To calculate the order of $C$, note that 
\[C=\langle \hat{F}_1, \mathbf{u}_p\hat{F}_2 \rangle = (\hat{F}_1) \oplus (\mathbf{u}_p\hat{F}_2). \]
Hence, $\lvert C \rvert = (p^2)^{2(n-deg\hat{F}_1)}(p^2)^{n-deg\hat{F}_2}=p^{2s}.$
\Q\Box
\section{Left $\FFF_p$-module isometry}
Recall that $M_2(\FF_p)=\mathcal{F}_{p^2}+\mathbf{u}_p\mathcal{F}_{p^2}$ and $B_p=\{(\alpha a_1+\alpha b_1 \omega)+u(\beta a_1+\beta b_1 \omega)|\alpha\in \FF_p^\times, a_1,b_1,\beta \in \FF_p \}$ is a subset of $\FF_{p^2}+u\FF_{p^2}$ Consider the mapping $\Phi_p$ defined as 
\[\Phi_p: M_2(\FF_p) \longrightarrow \FF_{p^2}+u\FF_{p^2}\]
where
\[\Phi_p\Biggl[\begin{pmatrix} a&b\\
                             b-c&a-b-d)
\end{pmatrix}\Biggl]= (a+b\omega)+u(c+d\omega).\]
It is easy to show that $\Phi_p$ is a left $\FF_p$-module isomorphism.
Now let $\mathcal{D}_p$ be the set of matrices in $M_2(\FF_p)$ with entries $a=\alpha a_1$, $b=\alpha b_1$, $c=\beta a_1$ and $d=\beta b_1$ where $\alpha \in \FF_p^\times$ and $a_1,b_1,\beta \in \FF_p$ then $\Phi_p^{-1}(\mathcal{B}_p\backslash\{0\})=GL(2,p)$. Therefore, $\Phi_p$ is a left $\FF_p$- module isometry such that $w_{\tt B}(A)=w_{\tt L}(\Phi_p(A))$ for all $A\in M_2(\FF_p)$. Thus, if $C$ is a cyclic code over $M_2(\FF_p)$ with minimum Bachoc distance $d_{\tt B}(C)$, the image $\Phi_p(C)$ is an additive cyclic code over $\FF_{p^2}+u\FF_{p^2}$, $u^2=0$ with minimum Lee distance $d_{\tt L}(\Phi_p(C))=d_{\tt B}(C)$.
\section{Examples}
For the following examples, MAGMA routines were created to construct cyclic codes over $M_2(\FF_p)$ and their isometric images. 
\begin{example}Let $p=2$ and $n=3$.  Then $x^3-\begin{pmatrix} 1&0\\ 0&1 \end{pmatrix}=F_0F_1F_2$ where $F_0=\begin{pmatrix} 1&0\\0&1 \end{pmatrix}$, $F_1=\begin{pmatrix} 0&1\\1&1 \end{pmatrix}$ and $F_2=\begin{pmatrix} 1&1\\1&0 \end{pmatrix}$.  Then $C_1=\langle\hat{F}_1,\mathbf{u}_2\hat{F}_2\rangle$ is cyclic code of length $3$ with $|C_1|=2^6=64$, minimum normalized homogeneous distance $d_{\tt nhom}=2$, minimum Bachoc distance $d_{\tt B}=3$ and minimum Hamming distance $d_{\tt Ham}=2$. The image $\Phi_2(C_1)$ is an additive cyclic code over $\FF_4+u\FF_4$ of length $3$, order $64$, and minimum Lee distance $d_{\tt L}=3$.
\end{example}

\begin{example} Let $p=3$ and $n=4$.  Then $x^4-\begin{pmatrix} 1&0 \\ 0&1 \end{pmatrix}= f_1f_2f_3f_4$ where $f_1=x-\begin{pmatrix} 1&0\\0&1 \end{pmatrix}$, $f_2=x+\begin{pmatrix} 1&0\\0&1 \end{pmatrix}$, $f_3=x+\begin{pmatrix} 2&1\\1&1 \end{pmatrix}$ and $f_4=x+\begin{pmatrix} 1&2\\2&2 \end{pmatrix}$.  If we let $F_0=f_2f_4$, $F_1=f_3$ and $F_2=f_1$ then $C_2=\langle \hat{F_1},\mathbf{u}_3\hat{F_2}\rangle$ is a cyclic code of length $4$ of order $|C_2|=9^3=729$ with minimum normalized homogeneous distance $d_{\tt nhom}=27/8$, minimum Bachoc distance $d_{\tt B}=4$ and minimum Hamming distance $d_{\tt Ham}=3.$ The image $\Phi_3(C_2)$ is an additive cyclic code over $\FF_9+u\FF_9$ with length $4$, cardinality $729$ and minimum Lee distance $d_{\tt L}=4$.
\end{example}

\end{document}